\documentstyle[11pt,dunk2001_asp,twoside]{article}
\def\edcomment#1{\iffalse\marginpar{\raggedright\sl#1\/}\else\relax\fi}
\reversemarginpar

\def\stacksymbols #1#2#3#4{\def\theguybelow{#2}
\def\verticalposition{\lower#3pt}
\def\spacingwithinsymbol{\baselineskip0pt\lineskip#4pt}
\mathrel{\mathpalette\intermediary#1}}
\def\intermediary#1#2{\verticalposition\vbox{\spacingwithinsymbol
\everycr={}\tabskip0pt
\halign{$\mathsurround0pt#1\hfil##\hfil$\crcr#2\crcr
\theguybelow\crcr}}}

\def\gapprox{\stacksymbols{>}{\sim}{4}{1}}

\begin{document}
\title{Aspects of Galaxy Formation}
 \author{Joseph Silk}
\affil{Department of Physics, Denys Wilkinson Building,
 Keble Road, Oxford OX1 3RH, United Kingdom}

\begin{abstract}
I describe some of the current challenges in galaxy formation theory
with applications to formation of disks and of spheroids.
Forthcoming deep surveys of galaxies with Keck and VLT 
will provide high
quality spectra of $\sim 10^5$ galaxies that will probe stellar populations
and star formation rates at redshift unity. This will help refine our
phenomenological knowledge of galaxy evolution and  enable robust
predictions to be developed for future breakthroughs in 
understanding galaxy formation  at high redshift that are anticipated with
NGST and with  the proposed new generation of 30 metre-class telescopes.

\end{abstract}

\section{Introduction}
Galaxy formation is a complex process, involving both gravity and
hydrodynamics, and can be complicated by such ingredients as turbulence and
astrochemistry. The disks and the spheroids of galaxies have undergone
distinct. although not necessarily uncoupled, histories.
The physics of disk formation has made considerable progress, in no small
part due to the pioneering review by Ken Freeman in {\it Stars and Stellar
Systems, Volume IX} that assembled diverse observational and theoretical  aspects
together for the first time. Spheroid formation is in a less satisfactory  state, in
part because spheroids are old and so their formation occurred long ago, and
there are correspondingly few direct clues. We do not yet have an adequate
understanding of either disk or spheroid formation. 

There is a simple reason for this predicament.
We have  no fundamental theory of star formation: the best we can do 
even in nearby regions of star fomation is to assemble
phenomenologically-motivated arguments and laws. When phenomenology is sparse as in the early universe, all bets are off as to the scalability  of current epoch theory to the past.  Of course, the lack of a robust theory 
has never deterred
theorists, and in this talk I will highlight some of the key
issues currently confronting cosmologists.

\section{The efficiency of star formation}

Textbooks state that disks are blue and bulges are red. The colours reflect the current star
formation histories of these diverse systems. Spiral galaxies  are
undergoing  star formation  at a
healthy rate some 10 Gyr or more after the disks formed, whereas 
spheroids such as that of our own galaxy have
long since (at least 5 Gyr ago) exhausted their gas supplies.
Reality is somewhat different, and there is no hard and fast discrimination
via colours between disks and spheroids.
It takes the merest trickle of star formation to bluen ellipticals
or to generate
stellar   population spectral line indicators symptomatic of relative youth.
Such objects are found  with increasing frequency as deeper and more complete surveys are performed. Of course, red disks are a characteristic of S0 galaxies.

\subsection{Disks}
Cold disks are gravitationally unstable,
and the instabilities are responsible for the formation of the giant
molecular clouds within which most stars form.
The cold gas concentration increases as the cloud
velocity dispersion is reduced. For a disk geometry,
both effects drive the Toomre gravitational
instability parameter $Q\propto\sigma_g/\mu$ down, 
where $\sigma_g$ is the disk velocity dispersion and $\mu$ is the disk  mass surface density.
Then
$Q\gapprox 1$ is the condition for the disk to be locally stable
against axisymmetric gravitational instabilities. This is also a necessary condition for
global stability against non-axisymmetric instabilities.  Lowering $Q$ 
further destabilizes the
disk, and increases  the star formation rate via cloud-collision induced star
formation.  

 In order for disks to be actively forming stars today, the efficiency of star
formation must be low in order for the initial gas supply not to have been
exhausted.
The present gas accretion rate onto the disk
as inferred from observations of the high velocity clouds    is too low by an order of
magnitude to sustain ongoing star formation. Indeed 
some of the high velocity clouds
are likely to be gas ejected from the galaxy rather than primordial clouds
sustaining a halo gas reservoir, because of their near-solar chemical
abundances.
At least one large high velocity cloud complex is dust-poor and 
metal-poor (Richter et al. 2001),
suggestive of primordial infall that has mixed with gas ejected from the disk.

A simple argument  for the low efficiency of star formation in disks appeals
to feedback from supernovae.
Let $v_{SN}$ be the specific momentum injected by supernovae per unit star
formation rate $\dot\rho_*$,  given by 
\begin{equation}
v_{SN} = E_{SN} /v_c m_{SN} \ = 500 E_{51}^{ 13/14} n_g^{- 1/7} m_{250}
\zeta_g^{-3/14}  \rm km\,s^{-1} \label{eq:1}
\end{equation}
with $v_c = 413 E^{1/4}_{51} n^{1/7}_g
\zeta^{3/14}_g$ being the velocity at which the remnant enters
the momentum conserving regime in a uniform disk of density $n_g$, $E_{51} \equiv E_{SN} / 10^{51}$ergs the
supernova energy (taken to be $10^{51}$ ergs), and $\zeta_g$ the
metallicity relative to solar of the ambient gas 
for an analytic fit to a spherically symmetric supernova remnant (Cioffi, Mckee and
Bertschinger 1988). Here $m_{SN}$ is the
mean mass required in forming stars in order to produce a supernova. For
SNII,
one simply assumes an initial mass function (IMF) with all stars of mass
above 8 M$_{\odot}$ becoming supernovae, so that $m_{SN} \approx 200$
M$_{\odot}$ 
for a Miller-Scalo IMF.
For a global star formation rate of $\sim 3 \rm M_\odot\,yr^{-1},$ the
inferred supernova rate is $\sim 1/70\, \rm yr$
for Type Ib, Ic and II supernovae (Capellaro et al. 1997).
 One can increase the inferred rate
by $\sim 50\%$ to include Type Ia supernovae
for an estimate of the total rate of  supernovae after the first
$10^8$ years have elapsed (to allow sufficient time for SNIa to form).

The momentum input from supernovae is dissipated via cloud-cloud
collisions and outflow from the disk. In a steady state, the momentum
input rate $\dot{\rho}_* v_{SN}$ must balance the cloud collisional
dissipational rate $\sim  p_g l^{-1}_t$ and the momentum
carried out in outflows $\sim  p_g H^{-1}$, where $p_g$ is the turbulent
pressure $\rho_g \sigma_g^2$ of the two-phase interstellar medium, 
$l_t$ is the cloud mean free path, given by $l_t\approx \sigma_g  \Omega^{-1}$ if  the cloud velocity dispersion is
induced by disk gravitational instabilities (Gammie,
Ostriker and Jog 1991),
$H
\equiv \sigma^2_g/2 \pi G \mu$ is the disk gas scale height, and $\mu$
is the surface mass density. 
Since $H\sim l_t,$ these two momentum dissipation rates are comparable.

The observed three-dimensional cloud velocity dispersion is 11 $\rm km\, s^{-
1}$ (for molecular clouds within 3 kpc of the sun) (Stark and Brand 1989).
I equate the star formation and star death rates,
and  model the star formation rate by initially only incorporating a dependence on local gas density and dynamical time: $\dot\rho_\ast=\epsilon\Omega\rho_g.$ Ignoring any outflow or
infall contributions to the momentum budget, one
balances turbulence generation by gravitational instability driven by
large-scale shear and differential rotation on large scales 
(Wada and Norman 1999)
with 
supernova momentum input on small scales (Silk 1987, Wada and Norman 2001).
A simple argument then  leads to 
\begin{equation}
\varepsilon \Omega \mu_g v_{SN} = \mu_g \sigma_g \Omega, \label{eq:2}
\end{equation}
 so that $\varepsilon = 0.02$ $(\sigma_g / 10 \rm km\, s^{-1}$) $(500 \rm
km\, s^{-1} / v_{SN})$. This reasoning
suggests that supernova feedback can indeed yield
the required low efficiency of star formation.

For a galaxy such as the Milky Way, the  global
star formation efficiency is expected to be around 2 percent, both as inferred from the global
values of gas mass $( \sim 6 \times 10^9 M_{\odot})$ and star formation
rate $( \sim 3 M_{\odot} $yr$^{-1}$) after allowance for gas return from
evolving stars (the returned fraction $\sim 0.5$ for a Miller-Scalo IMF)
over a galactic dynamical time and as more
directly inferred  from
studies of HII region radio luminosities summed over molecular cloud masses
 (e.g. Williams and McKee 1997).
Thus the Milky Way
interstellar medium has a
predicted efficiency of star formation  comparable to what is
observed. 

One consequence of a gravitationally unstable cloud-forming
and star-forming disk is that the turbulence seen today in   cloud motions
yields an effective viscosity that can account for various properties of
galactic disks,
including exponential surface brightness profiles
(Silk and Norman 1981; Lin and Pringle 1987), the disk scale size (Silk
 2001),
the molecular gas fraction (Vollmer and Beckert 2001),
the Tully-Fisher relation (Firmani and Avila-Reese 2000),
and the star formation rate and efficiency (Devriendt, Slyz and Silk 2002).

\subsection{Spheroids}

In contrast with galactic disks, star formation rates were once high and
efficient in spheroids, when they were gas-rich.
The obvious difference is geometry: the gas velocity dispersion, and hence
gas pressure, is much higher in forming spheroids. A more complex model is
needed for the interstellar medium that explicitly incorporates the
3-dimensional geometry  and the multiphase interstellar medium.

Even minor mergers  result in gas being driven in substantial amounts into
the central regions of the galaxy.
The stage is set  for spheroid formation. The gas mass and concentration is
so high that a starburst must surely develop, as indeed is observed.
But the detailed conversion of gas into stars is poorly understood.
Supernovae must play an important  role
in providing momentum feedback and thereby controlling the duration of the starburst.

I consider a two-phase medium in which dense cold clouds are embedded in the
hot, supernova-heated diffuse medium. I model the volume of the hot phase 
by porosity, $f=1-e^{-P}$, and argue that 
the porosity $P$ of the hot medium controls the stellar feedback.
The porosity, defined below,
 is a measure of
the fraction of volume $f$ occupied by the hot phase ($T
\sim 10^6\,{\rm K}$) associated with the interiors of
supernova remnants. 
In the context of disk formation and evolution,
breakout from the cold disk  occurs if the porosity is large,
so that the supernova-heated bubbles can penetrate into the halo, and most of the kinetic
energy injected by the supernovae flows out in chimneys or   fountains.
A plausible condition for self-regulation is $P\sim 0.5$.

If the porosity is large, outflows develop, and star formation is initially
enhanced by compression of cold clouds.  As the cold gas is depleted,
by both star formation and outflows, star formation eventually is quenched. In a disk geometry, the 
winds drive supernova ejecta out of the disk and thereby make feedback
ineffective. In a spheroid, it should be  easier to drive an outflow
through the diffuse medium if  $P$ is not too small.

It seems likely that the enhanced gas
concentration in the low $P$ limit will
drive up the star formation rate and initiate  a starburst. This at least is the
generic assumption that underpins virtually all studies of merger-induced
star formation.
In other words, low porosity enhances the feedback from supernovae, thereby
driving up the porosity. Hence $P\sim 0.5$ seems to be the natural outcome of
the resulting self-regulation of star formation, with a hot gas fraction
$f\sim 0.5$  applying in a quasi-steady state. 
With self-regulation, one is in the low efficiency regime.
Hence this will be the long-term fate of a starburst as the gas supply is
diminished, by  consumption in star formation  and by outflow.

Starbursts are usually considered to characterize massive spheroid formation.
Observations of ULIRGs certainly  imply high efficiency of star formation,
evidence for triggering by mergers at least in extreme cases,
and rapid generation of a de Vaucouleurs-like profile.
Nevertheless one persistent line of reasoning that stems from
the cold dark matter scenario for hierarchical galaxy formation has 
insistently and reasonably successfully argued that 
spheroids, apart from their nuclei, form from dissipationless mergers
of galaxies. 
The stars form before the spheroid is assembled.
In this way, one can have an old stellar population
in place by $z\sim 1$, where observational evidence seemingly
insists that only passive evolution has occurred for E and S0
galaxies in clusters and even in the field.
The occasional indications of intermediate age features (Balmer absorption lines etc.) seen especially in some field ellipticals are explained by very low rates of recent star formation (Ferreras and Silk 2000).

Hence low porosity leads to low feedback and high efficiency of star
formation, while high porosity means strong feedback and low efficiency.  Of
course if the porosity is too large $(P\gg1)$, any winds or outflows are
likely to be suppressed via superbubbles that overlap and self-destruct.  The
superbubble interiors are fed by evaporation of cold entrapped gas clouds,
and this is the source of the outflows. Hence it is logical to expect that
for $P\sim 1$, a wind is driven. Indeed observations of star-forming
galaxies, including starbursts, show that ouflow rates are on the order of
the star formation rate (Martin 1999; Heckman et al. 2000).

\section {An analytic approach to star formation rates and efficiency}

One may quantify these  arguments on porosity as follows.
Porosity is defined to be the product of the supernova remnant 4-volume at
maximum extent, when halted by ambient gas pressure, and the
rate of bubble production.  
  One can then write  the
porosity as  $P=(\dot\rho_* /m_{SN})
\left(\frac{4}{3} \pi\, R^3_a\, t_a \right),$
where $\dot\rho_*$ is the star formation rate per unit volume, $m_{SN}$ is
the mass in stars formed per supernova, and $R_a$ is the
radius of the supernova remnant at time $t_a$ when halted by
the ambient (turbulent) gas pressure $p_g$.  One finds that
the porosity $ P \propto \dot\rho_*\, p^{-1.36}_g \rho^{-0.11}_g $
is extremely sensitive to the interstellar pressure.

It is relevant to look at the porosity of nearby star-forming galaxies.  Oey,
Clarke and Massey (2001) note that one can approximate
 the porosity as $$P\approx
16\frac{\Psi(\rm M_\odot yr^{-1})}{h_dR_d^2(\rm kpc^3)},$$ where a Salpeter
IMF has been adopted and an ambient disk interstellar medium pressure
$p/k=9500\rm\,cm^{-3}K$ has been assumed.  Here $h_d$ and $R_d$ are the gas
disk scale-height and scale-length, respectively, and $\Psi$
is the global star formation rate. The Local Group galaxies
display a wide range of global porosities, from the extreme case of IC10
$(P\sim 20)$ to M33 $ (P\sim 0.3)$ and the SMC $(P\sim 0.2).$

 In fact, a
value of $P$ of order unity, as inferred both for the LMC and for the Milky Way (this
latter case being based on the observed supernova rate) would seem to be not
untypical for large galaxies $(\gapprox 0.1L_\ast)$, admittedly based on rather
poor statistics.
While the situation for  our own galaxy is confusing with regard to
direct HI mapping and determination of  $P$ (Heiles 2000), there is substantial infall to and outflow
from the galactic disk as seen in OVI surveys performed by the FUSE
satellite. One can make a strong case that
$P\sim 1$ for the LMC from HI maps and that there is substantial injection of
mechanical energy from regions of star formation
 into the diffuse interstellar medium by expanding HI
supershells (Kim et al. 1999).
The  OVI absorption studies show that
the mass-flow rate from one side of the LMC disk is
about $1\rm M_\odot yr^{-1}$ (Howk et al. 2001).
This is comparable to the global star formation rate for the LMC.

The role of supernovae in driving the observed superbubbles is
inferred  indirectly  but supernovae appear to provide the
dominant injection of energy. Excess expansion rates are measured relative to the standard assumptions for OB stellar wind-driven outflows, and 
excess x-ray luminosities are measured relative to the
estimated post-shock luminosities. The occurence of several supernovae within a given superbubble is a natural expectation given any reasonable IMF, and 
seems to be required by the observations.

 The porosity  ansatz
 provides the motivation for the feedback
prescription.
By incorporating an analytic fit to the
evolution of a spherically symmetric supernova-driven shell, one can
write
\begin{equation}
P = G^{-\frac{1}{2}} \sigma^{2.72}_f p^{-1.36}_g \rho^{-0.11}_g
\dot{\rho}_* \ , \label{eq:3}
\end{equation}
where $\dot{\rho}_*$ is the star formation rate, $p_g$ is the ambient
gas pressure, both thermal and turbulent, and $\sigma_f$ is a fiducial
velocity dispersion that is proportional to $E^{1.27}_{SN} m^{-1}_{SN}
\zeta^{-0.2}_g$ and may be taken to be 18  $ \rm km \, s{^{-1}}$ for $E_{SN} =
10^{51}$ erg, $m_{SN} = 200$ M$_{\odot}$ and $\zeta_g = 1$. Note that at
large pressure, $P \ll 1$ and porosity is primarily controlled by the ambient
pressure, which I take to be dominated by turbulence: $p_g = \rho_g
\sigma^2_g$.

 Consider the
possibility of strong feedback. In this case,  the filling factor of hot
gas is of order fifty percent, which is the requirement for strong feedback.
Rewriting (3) as 
\begin{equation}
\dot{\rho}_* \approx G^{1/2} \rho^{3/2}_g \left(
\frac{\sigma_g}{\sigma_f} \right)^{2.7} P \ ,  \label{eq:4}
\end{equation}
one explicitly incorporates feedback into the star formation rate.
The
star formation rate is controlled by
ambient pressure. Thus even in starbursts, as long as $P\sim 0.5$, a
Schmidt-type relation is maintained. The
predicted efficiency is around 2\% for the Milky Way disk, in agreement with
the  estimate from
(2), but 
can be as high as 50\%, for merger-induced turbulence ($\sigma_g \sim 50\, \rm
km\, s^{-1}$) and a 
standard IMF.

If the turbulence is low, the efficiency is much less.
However the porosity may still be large
so that feedback is strong.
Inserting the derived expression (2) for 
disk star formation efficiency into the
equation for the porosity, I find that
the porosity can be inferred: 
$$ P \approx \left( \frac{\rho}{ \rho_g} \right)^{\frac{1}{2}} 
\left(\frac{\sigma_f}{ \sigma_g} \right)^{1.7}
\left( \frac{\sigma_f}{v_{SN}}\right) 
= 0.5 \left( \frac{\rho / \rho_g}{0.1}
\right)^{\frac{1}{2}}
\left( \frac{\sigma_g}{10 \rm{km\, s}^{-1}} \right)^{- 1.7}.$$

The feedback is weak $(P\ll 1)$  if the  gas pressure is large.
Very high turbulence quenches the porosity, because of the high gas pressure.
In a starburst, the star formation rate is high and  the efficiency is high, but the porosity of the hot medium is initially low.  
Feedback is small and one has runaway star formation.
Nothing impedes gas accretion, cooling and collapse.
The low  porosity does not necessarily quench outflows
which can be carried by the neutral gas, and hence driven by the 
momentum input into the neutral interstellar medium.
The runaway star formation will result in an increase in the porosity.
Hence $P\sim 1$ should apply most of the time.
Outflows will then be common. 

 A top-heavy IMF
could substantially
reduce $v_{SN}$, and star formation efficiences of order 50\% would then
readily be attainable, even with modest levels of interstellar turbulence.
Such efficiencies may be needed in order to account
for the luminosities measured in some  ultraluminous infrared galaxies where the
molecular gas masses are measured.  A top-heavy
stellar initial mass function might be required in protogalaxy mergers in
order to reconcile the hypothesis that ellipticals formed in such events with
the inferred paucity of young ellipticals at intermediate redshifts.

There may be other indications that the IMF  in the early universe may be more weighted to massive stars and light production than the local IMF.
A   quantitative comparison is between the star formation rate and measured
rest-UV luminosity density from star-forming galaxies at $z\sim 3$ with the local $K$-band luminosity density, where, for reasonable extinction corrections, 
a possible overproduction of old starlight is inferred for a local 
near infrared IMF (Cole et al. 2001). Similar conclusions come from diffuse extragalactic background light. Observations at optical wavelengths 
 (Bernstein, Freedman and Madore 2001)
combined with the UV and FIR diffuse extragalactic light backgrounds yield a total
extragalactic background light density of $100 \pm 20 \rm nw\, m^{-2}\,sr^{-1}.$
Such an intensity, predominantly from high redshift galaxies, overproduces the local stellar density by about a factor of 2 for a standard IMF.

\section{A unified approach to galaxy formation }

There are three approaches that have been explored towards developing a unified approach to galaxy formation.
\subsection {Numerical hydrodynamics}
The loss of protogalactic angular momentum is confirmed by high resolution
simulations.
Disks are a factor $\sim 10$ too small. The resolution of this problem
requires more realistic modelling of disk formation that incorporates gas
physics and stellar feedback.

Disks are two-dimensional systems, which 
makes their stability easier to model. The three-dimensional components
of galaxies are not well understood. One has made most progress with the the
dark matter halos, although the 
characteristic halo scale defined by the density profile
is also controversial. The NFW profile
(Navarro, Frenk and White 1997) certainly has a scale,
 but it may not be the correct
scale as defined by the dark matter cores
according to high resolution simulations of dark matter halos.
The concentration of dark matter  appears to be   excessive in the inner
disk of our galaxy (Binney and Evans 2001)
and in barred spirals (Debattista and Sellwood 2000).
Central dark matter cusps are predicted  (Ghigna et
al. 2000; Jing and Suto 2000;
Klypin et al. 2001) that are  not observed in LSB dwarfs (van den Bosch and
Swaters 2001). The halo  substructure results in considerable
angular momentum losss from the dissipating baryons to the dark matter,
forming 
disks that are far too small (Navarro and Steinmetz 2000).
All of these problems presently plague numerical modelling of disk galaxy formation.

These various difficulties on subgalactic scales have been touted as creating a crisis for
the cold dark matter scenario of galaxy formation, hitherto so remarkably successful on larger scales.
Possible  resolutions
 come under two distinct guises: tinkering with the
particle physics or elaborating on the astrophysics.
The former class of solutions  appeals  to invoking new theories of gravity that may even dispense entirely with the need for any dark
matter, or  to
 the introduction of exotic varieties  of
particle  dark matter, such as self-interacting or warm dark
matter. The astrophysical possibilities include various types of feedback,
that might involve dynamical interactions between baryons and the dark matter.
Proposals include a  combination of angular momentum transfer to and 
heating of dark matter  via  a massive primordial rotating  baryonic  bar 
which would undergo 
 resonant interactions
with the dark matter (Weinberg and Katz 2001)
and   drive  massive gaseous 
 outflows (Binney, Gerhard and Silk 2001).

Bulge sizes are equally a mystery, as far as any fundamental theory of galaxy
formation is concerned.
At least one may hope that the scale of disks ultimately comes 
from angular momentum considerations, with the initial angular momentum
being obtained from second-order theory of tidal torques between density fluctuations in combination with feedback considerations. 
The final bulge sizes are determined by the complex star-gas interactions in a starburst.

Fully numerical treatments of galaxy formation
include N-body simulations to follow the dark matter and the stars, coupled with hydrodynamics to follow the gas dynamics.
The procedure has succeeded in commencing with cosmological scales, 
zooming in at progressively higher resolution with adaptive mesh techniques
and a grid-based code to resolve the scales on which the first stars may have formed, around $100\rm M_\odot$ (Abel, Bryan and Norman 2000).\cite{abe}
This at least is where the fragmentation
seems to terminate. The resulting gas clumps are identified with 
the first stars.

Unfortunately the adaptive approach only succeeds in resolving a single clump
that is assumed to be repesentative of the entire protogalaxy.  Moreover,
there is no guarantee that further fragmentation does not occur. Additional
complexities include feedback from the first massive stars that will modify
subsequent cooling, fragmentation and star formation. The numerical scheme of
adaptive mesh refinement cannot yet tackle global aspects of galaxy
formation. This approach is ideal however for providing the crucial subgrid
physics which one can eventually hope to combine with galaxy-scale
simulations.

Another approach to modelling feedback is to use smooth particle hydrodynamics (SPH)
to model a large volume of the universe, resimulating the dense peaks where
galaxies form at high resolution.
SPH allows feedback to be studied  over galaxy scales.
However,
feedback prescriptions have not hitherto been effective or convincing, forming either forming disks too late by arguing that
feedback globally delays gas cooling and hence  star formation
(Weil, Eke and Efstathiou 1998), or  by
inserting an ad hoc delay between energy feedback from massive star deaths and gas  cooling (Couchman and Thacker 2001).

Sommer-Larsen (2002) has  incorporated simple star formation rules
into a feedback model, and has succeeded with a SPH code in producing disks
that are within a factor 2 of the observed size.
The disks form inside-out, but still have a substantial number of old stars
from previous accretion events in their outer parts,
as seen in the outer regions of M31 (Ferguson and Johnson 2001).
Subsequent gas infall occurs even at the current epoch, and x-ray
observations of halos provide  an important constraint on models of disks and
spheroids. Isolated ellipticals  in particular should contain a
considerable  reservoir of
hot gas.

\subsection {Semi-analytical galaxy formation}

Semi-analytical theory  uses N-body simulations to sample a large volume and determine the density peaks and velocity minima
where galaxy formation is likely to occur.
 Monte-Carlo realisations  are constructed of the merging histories
of dark matter halos.
Continuing cooling occurs around dense peaks in the density field. Cooling is
rapid  in these regions, and this is where the galaxies form.
Gas disks are the basic objects to form first.
Mergers occur in the denser regions where there are adjacent peaks.
These are the regions that eventually form clusters of galaxies.
More quiescent accretion occurs in the relatively isolated regions.

Star formation rules are then applied in each local environment
to generate galactic disks 
in the accretion-dominated regions via a Schmidt-type law.
Ellipticals form 
in the merging environments  via starbursts.
There is no effective resolution of galaxy scales.
Initial scales  are associated with the radius at maximum extent of an overdensity of specified mass, with  angular momentum 
assumed to subsequently be 
conserved. The protogalactic cloud forms with the small value of initial
dimensionless angular momentum $\lambda \approx 0.05$ that is generated by tidal torques
between neighbouring fluctuations, and attains virial equilibrium at a fraction $\lambda$
of the maximum radius.
A simple  law is used for the star formation rate
in the resulting disk, taken to be proportional to the gas density divided by the local free fall time,
and cold gas infall continues to feed the disk.

Population synthesis combined with dust modelling yields colours and counts
(Somerville \& Primack 1999; Benson et al. 2000;
Devriendt and Guiderdoni 2000; Kauffmann \& Haehnelt 2000). Spiral galaxies are relatively isolated galaxies that have not undergone a significant
merger  within the past 5-10 Gyr.  Elliptical galaxies are assumed to form via major
mergers on a dynamical time scale, with  stars  assumed to form in a starburst.

Once the parameters are carefully adjusted, including feedback, 
the semi-analytical approach provides good agreement with the observed
luminosity function, multi-wavelength band galaxy counts, redshift distributions and cosmic star formation histories. It has successfully reproduced the clustering
of the Lyman break galaxy populations.
Accretion occurs primarily along filaments and sheets of dark matter.
 Early galaxy formation occurs in overdense filaments of dark matter. 

The theory is less predictive on issues that involve star formation. In the
hierarchical theory, most stars form relatively late, and it is not clear
whether the colours, spectra and luminosities of galaxies at redshift unity
can be reproduced. 
Forthcoming surveys will provide the data base with which the current models
can be definitively tested and refined.
Unfortunately, the semi-analytic models are not especially robust. There are
several adjustable parameters, and 
detailed exploration of the large parameter space for the gas
 and star formation physics is not  feasible.
Nor is there any resolution of the question of the determination of disk or
bulge sizes. Ultra-high resolution simulations will be needed in order to
make
substantive progress in our understanding of how galaxies formed.

\subsection {Phenomenological  galaxy formation}

Analytical theory  comes in different flavours, within a strongly
phenomenological context. Backwards galaxy formation is one example, where models are made of nearby disks that are evolved back in time.
This has the advantage of allowing the incorporation of realistic 
star formation modelling, but cannot easily cope with mergers and infall.
Angular momentum conservation,
not necessarily the best of approximations, plays a key role in  generating
low surface brightness galaxies (Dalcanton, Spergel and Summers 1997)
and in modelling the Tully-Fisher relation  for disk galaxies (Mo, Mao
and White 1998). Accretion of gas and viscous disk self-regulation are central to the approach of Firmani, Hernandez and Gallagher (2000).

A phenomenological variant has been developed that involves galaxy collisions and mergers
(Balland,  Silk \& Schaeffer 1998) by using tidal interactions to determine where and when the different morphological types form, 
normalizing  the  theory to the morphological dependence on local density.
This approach is almost completely phenomenological. 
 Ellipticals form by rare major mergers and  disks form by
prolonged infall, which is equivalent to a sequence of minor mergers.

The analytic approach to merger  physics is parametrised by an approximate fit to the collision
cross-section to estimate the energy exchange 
incurred in tidal
interactions and mergers.  Strong interactions leading to mergers are assumed
to form ellipticals, with intermediate strength interactions between galaxies
generating harassment (Moore
et al.  1999) that results in S0 formation.  Minor interactions,
typical of the field, result in spiral galaxy formation.  The fraction of
elliptical, S0 and spiral galaxies can
 be predicted as a function of redshift and of mean local density.

To make contact with observations, star formation and chemical 
evolution must be incorporated. 
Gas cools radiatively within the halos,
settles into a disk and forms stars. Semi-empirical recipes are
used to account for various astrophysical processes,
including  star formation efficiency,
dust opacity, absorption and emission, and  feedback
(Devriendt and Guiderdoni 2000).
The  spectral evolution model  (Devriendt, Guiderdoni \& Sadat 1999) self-consistently links the optical and the far-IR/submillimeter emission.
One  then selects 
every galaxy identified as an early 
type galaxy to undergo an ``obscured starburst'' phase, whose intensity and duration are controlled by the amount of gas available for star formation. 

A fair overall agreement is found between between models and data
(Silk and Devriendt 2001; Balland, Devriendt and Silk 2002).
Late-type galaxies dominate the counts and background light relative to early types 
in the optical and the far-IR,
but at longer wavelengths the contribution
of early-type galaxies  exceeds 
 that from late types. 
This behaviour is of course due to the  negative
k-correction, which makes galaxies of the same bolometric luminosity
as bright 
at redshift 5 as at redshift 0.5.
This effect is only important in the submillimeter (e.g. for SCUBA at 850 
microns), because the peak  rest frame 
emissivity of dust occurs between 60 and 100 microns.
Since the
 S0s and ellipticals form at $z > 2-3$, the associated  redshifted
emission dominates the diffuse background  at wavelengths
greater than about 300 microns.

\section{ Some closing remarks}

Stellar evolution, from birth to death, is   the key to understanding galaxy
formation. Observations of galaxy evolution are flourishing as never before,
thanks to the availability of the 8 metre-class telescopes.
Imminent surveys with Keck and  VLT  will
provide samples of $\sim 10^5$ galaxies at $z\sim 1.5$  with sufficient
spectral resolution to study stellar population and star formation rate evolution. 

Theory lags far behind the data.
Our best hope may be to construct a phenomenological model that
incorporates the successes of the numerical simulations and of the
semianalytic studies. This will surely involve
a backwards approach,
using the nearby universe to effectively
simulate the universe at redshift unity.
This has already been done, apart from the essential complication 
now under intensive study of
developing the small-scale (stellar
and interstellar) physics input that is so crucial for 
understanding and modelling feedback.

Refining this model with the new data sets 
that we anticipate from the DEIMOS and VIRMOS surveys,
we can then hope to take the next step backwards in time by developing
predictions for NGST and the 30 metre-class telescopes thare now under design study to probe
the epoch of the first galaxies, at $z\sim 6,$
when reionization occurred.
The future beckons brightly.

\end{document}